\documentclass[journal]{IEEEtran}

\usepackage[utf8]{inputenc}
\usepackage[T1]{fontenc}
\usepackage{amsmath,amssymb,amsfonts,amsthm}
\usepackage{graphicx}
\usepackage{textcomp}
\usepackage{xcolor}
\usepackage[sort,compress]{cite}
\usepackage{hyperref}
\usepackage{orcidlink}
\usepackage{microtype}
\usepackage{booktabs}
\usepackage{makecell}
\usepackage{threeparttable}
\usepackage{siunitx}
\usepackage{tikz}
\usetikzlibrary{decorations.pathreplacing, backgrounds, fit, positioning, calc, arrows.meta}

\newif\ifrevhighlight
\revhighlightfalse
\newcommand{\rev}[1]{%
  \ifrevhighlight\ifvmode\leavevmode\fi\begingroup\color{blue}#1\endgroup\else#1\fi}

\graphicspath{{./figures/}}

\hypersetup{
  hidelinks
}

\makeatletter
\def\subsubsection{\@startsection{subsubsection}{3}{\parindent}%
  {1.0ex plus 0.5ex minus 0.5ex}{0ex}{\normalfont\normalsize\itshape}}
\makeatother

\begin{document}

\title{SmartBAN on Silicon \\
  by Structured Behavioral Modeling}

\author{
  Masato~Yoshimi~\orcidlink{0000-0003-1343-9795},~\IEEEmembership{Student~Member,~IEEE,}
  Takahiro~Ito~\orcidlink{0000-0001-9796-9296},~\IEEEmembership{Member,~IEEE,} \\
  Kento~Tanaka~\orcidlink{0000-0002-3532-6954},
  and Hirokazu~Tanaka~\orcidlink{0000-0002-4647-9075},~\IEEEmembership{Senior~Member,~IEEE}%

  \thanks{\rev{This research and development work was supported by the MIC/FORWARD \# JPMI250830001.} (Corresponding author: Masato Yoshimi.)}%

  \thanks{M. Yoshimi, T. Ito, and H. Tanaka are with Graduate School of Information Sciences, Hiroshima City University, Hiroshima, Japan \rev{(e-mail: myoshimi@e.hiroshima-cu.ac.jp)}.}%

  \thanks{\rev{K. Tanaka is with Strategic Technology Center, Technology SBU, TISI Inc., Tokyo, Japan.}}%

  \thanks{\rev{H. Tanaka is also with Graduate School of Informatics, Osaka Metropolitan University, Osaka, Japan.}}%

  \thanks{\copyright~2026 IEEE. Personal use of this material is permitted. Permission from IEEE must be obtained for all other uses, in any current or future media, including reprinting/republishing this material for advertising or promotional purposes, creating new collective works, for resale or redistribution to servers or lists, or reuse of any copyrighted component of this work in other works.}%

  \thanks{Accepted for publication in IEEE Internet of Things Journal, DOI: \href{https://doi.org/10.1109/JIOT.2026.3747296}{10.1109/JIOT.2026.3747296}.}
}

\markboth{ACCEPTED FOR PUBLICATION IN IEEE INTERNET OF THINGS JOURNAL}%
{Yoshimi \MakeLowercase{\textit{et al.}}: SmartBAN on Silicon}

\maketitle

\begin{abstract}
  \rev{Wireless body area networks (WBANs) underpin the Internet of
Medical Things (IoMT).
IEC/ETSI SmartBAN defines a lightweight
time-division multiple access (TDMA) protocol, yet published
implementations on commercial hardware remain partial.
The standard specifies frame formats but leaves device-internal
behavior in phase control, connection lifecycle, and slot timing
unspecified.
We present an operational SmartBAN realization on a
commercial wireless system-on-chip (Nordic nRF54L15, Arm Cortex-M33,
Zephyr RTOS) that implements initial connection, scheduled uplink, and
timeout-based disconnection detection through a four-layer software
architecture whose hardware abstraction layer is the intended
portability boundary.
Hub and Node Mealy automata supply one concrete design for the
unspecified phase control and connection lifecycle, as a blueprint for
the firmware's event-driven dispatch.
Validation with 16 concurrent Nodes over 25~hours
shows conformance across every transition the build exercises (all but
one), ${>}$99.99\% packet delivery, and timeout-triggered reconnection
after the single observed disconnection.
An analytical current model fits the measurements once it uses measured
state currents; with datasheet currents it underestimates the average,
mostly in the radio-off state, whose current we attribute to
implementation-level overhead rather than the hardware.
A same-SoC comparison with Bluetooth Low Energy (BLE) shows how each
protocol departs from its schedule: SmartBAN, configured to send a fresh
sample every interval instead of retransmitting, kept each Node's frame
spacing within \qty{1.1}{\milli\second} of the interval but lost some
frames; BLE lost none but deferred some by one or two intervals, and drew
less than half SmartBAN's average current.}

\end{abstract}


\section{Introduction}

\IEEEPARstart{C}{ontinuous} physiological monitoring through body-worn
sensors underpins the Internet of Medical Things
(IoMT)~\cite{Movassaghi2014WBAN, Zhong2022Survey}.
As applications span clinical diagnostics and everyday wellness,
wireless body area networks (WBANs) must deliver low latency,
low power, multi-node support, and reliability under body
shadowing.

To address these requirements, SmartBAN~\cite{ETSI_TS_103_325, ETSI_TS_103_326, IEC_63203_801_1, IEC_63203_801_2}
was standardized as
a lightweight WBAN protocol whose time-division multiple access
(TDMA)-based channel access~\cite{Paso2015} \rev{assigns each sensor node
a dedicated uplink slot}.
While Bluetooth Low Energy (BLE)~\cite{BluetoothCoreSpec, Gomez2012}
dominates commercial wearables, its connection-event scheduling\rev{~\cite{Leonardi2018} gives no
  bound on when a sample arrives: a notification that misses its event is
deferred to the next one}~\cite{Kindt2020}.\footnote{\rev{Contention
    between peripherals is one cause of such deferrals~\cite{Kindt2020}.
    Our measurements do not isolate the mechanism, so we take it from the
    literature.}}
Despite mature specifications covering physical layer (PHY) and media
access control (MAC) layers, \rev{neither SmartBAN nor
  IEEE~802.15.6~\cite{IEEE802.15.6:2012} specifies device-internal
  behavior in enough detail to be implemented directly, and published
  implementations on commercial off-the-shelf wireless
systems-on-chip (SoCs) remain partial}; prior work is reviewed in Section~\ref{sec:standard}.
\rev{This} specification-to-implementation gap may have limited SmartBAN
adoption relative to BLE.

\rev{We bridge this gap through structured behavioral modeling.
  Formulating the unspecified behaviors as finite automata provides
  a design blueprint expressed as state transitions.}\footnote{\rev{Our group's earlier SmartBAN work, on TI~CC2650
    prototypes~\cite{Hiramatsu2023URSI, Hiramatsu2024BodyNets, Ito2024ISMICT,
    Yoshimi2024GCCE} and on the nRF54L15~\cite{Yoshimi2026EMBC}, is discussed
    in Section~\ref{sec:related}.}}
The main contributions are:
\begin{enumerate}
  \item \rev{\emph{A SmartBAN implementation on a commercial wireless SoC},
    realizing the baseline initial connection and scheduled uplink
    operations and timeout-based disconnection detection via a layered architecture (Section~\ref{sec:implementation}).}
  \item \rev{\emph{A structured behavioral model}: two Mealy-type finite automata
    that give the phase-control and connection-lifecycle gaps (Gaps~1 and~3,
    Section~\ref{sec:gaps}) one concrete design, a blueprint that maps to event-driven
    firmware dispatch (Sections~\ref{sec:design}--\ref{sec:implementation}).
    The implementation fills the platform-dependent Gaps~2 and~4.}
  \item \rev{\emph{Multi-axis experimental validation} with 16
    concurrent sensor nodes---evaluating behavioral conformance,
    communication quality, 25-hour stability, and operation under
    controlled RF interference---plus a unified same-SoC BLE comparison
    of delivery timing and energy consumption
    (Sections~\ref{sec:experiments}--\ref{sec:discussion}).}
\end{enumerate}

\section{SmartBAN Specifications and Design Gaps}
\label{sec:standard}

\subsection{Protocol Overview}
\label{sec:protocol-overview}

SmartBAN is a WBAN protocol
standardized by ETSI~\cite{ETSI_TS_103_325, ETSI_TS_103_326} and later adopted
as IEC~63203-801~\cite{IEC_63203_801_1, IEC_63203_801_2}. It defines a
star-topology network in which a single Hub coordinates up to 16~sensor
Nodes~\cite{Paso2015}.

The PHY~\cite{ETSI_TS_103_326} defines 40~channels at
\qty{2}{\mega\hertz} spacing in the \qty{2.4}{\giga\hertz} ISM band
(2402--\qty{2480}{\mega\hertz}), using Gaussian frequency-shift keying
(GFSK) modulation at \qty{1}{\mega\bit\per\second} with optional
Bose--Chaudhuri--Hocquenghem~(BCH) forward error correction.
Three of these channels serve as control channels~(CCH) for
frequency-diverse beacon advertisement; a single
data channel~(DCH) carries scheduled traffic.

The MAC~\cite{ETSI_TS_103_325} structures communication into a repeating
inter-beacon interval~(IBI). Each IBI comprises a D-Beacon carrying
a timestamp and slot-allocation map, a scheduled access \rev{period}~(SAP)
of deterministic TDMA uplink slots with immediate
acknowledgment~(I-Ack), and a contention management period~(CMP)
where unconnected Nodes send connection requests via slotted Aloha.
\rev{This structure allocates each connected Node at least one dedicated uplink
slot per~IBI.}

Viewing the MAC specification~\cite{ETSI_TS_103_325} from a
behavioral perspective yields six principal operations summarized
in Table~\ref{tab:mac_ops}\rev{, whose Coverage column marks which of them this paper implements, in full or in part}.

\begin{table}[!t]
  \caption{Six Principal MAC Operations of SmartBAN (Clauses per ETSI TS~103~325~\cite{ETSI_TS_103_325})}
  \label{tab:mac_ops}
  \centering
  \setlength{\tabcolsep}{5pt}
  \renewcommand{\arraystretch}{1.05}
  \footnotesize
  \begin{tabular}{@{}clcl@{}}
    \toprule
    \textbf{ID} & \textbf{Operation} & \textbf{Clause} & \textbf{Coverage} \\
    \midrule
    (1)  & Initial connection                      & 7.2.2        & This paper \\
    (2a) & Uplink (Scheduled Access Period)        & 7.3.1        & This paper \\
    (2b) & Uplink (Contention Management Period)   & 7.3.2        & \rev{Extension} \\
    (3)  & Supplementary downlink                  & 7.4          & Extension  \\
    (4)  & Slot re-assignment                      & 7.5          & Extension  \\
    (5)  & Data channel migration                  & 7.6          & Extension  \\
    (6)  & Disconnection                           & 6.2.6, 6.2.7 & \rev{Partial} \\
    \bottomrule
  \end{tabular}
\end{table}

\subsection{Specification Gaps}
\label{sec:gaps}

While the standard~\cite{ETSI_TS_103_325, ETSI_TS_103_326} defines
frame formats, channel parameters, and the IBI structure, it leaves
four categories of internal device behavior unspecified:
\begin{enumerate}
  \item \textbf{Phase control.} The standard distinguishes CCH for
        beacon advertisement and DCH for scheduled communication, but
        the internal mechanism for switching between them---CCH entry,
        network re-acquisition after beacon loss, and DCH resumption
        upon C-Beacon reception---remains unspecified.
  \item \textbf{Slot-level timing.} The standard defines slot
        boundaries and IBI duration yet provides no guidance on
        how Nodes derive per-slot radio timing from D-Beacon
        timestamps or compensate for inter-IBI clock drift.
  \item \textbf{Connection lifecycle.} Connection-related frame types
        (C-Request, C-Assignment, I-Ack) and endpoint states
        (\emph{connected}, \emph{unconnected}) are named, but the
        complete transition logic---triggers, guards, timeout-based
        disconnection detection, and autonomous recovery
        procedures---is absent from the specifications.
  \item \textbf{Scheduling policy.} Although the Hub assigns slots
        via C-Assignment, the standard does not prescribe the
        allocation algorithm---fairness criteria, priority handling,
        or dynamic reallocation upon node join/leave events.
\end{enumerate}
\rev{These gaps cover the behaviors essential for minimal star-topology operation.
Structured behavioral modeling (Section~\ref{sec:design}) gives phase
control (Gap~1) and connection lifecycle (Gap~3) one concrete design, a blueprint for
event-driven firmware. Slot timing (Gap~2) and scheduling policy (Gap~4)
depend on platform parameters, so the implementation
(Section~\ref{sec:implementation}) chooses them.}

\subsection{Related Work}
\label{sec:related}

\rev{Table~\ref{tab:related} places this work among prior WBAN and
SmartBAN studies.}
IEEE~802.15.6~\cite{IEEE802.15.6:2012} realizations remain limited to
partial PHY-layer prototypes on FPGA or SDR
platforms~\cite{Sayed2015FPGA, Su2017Prototype}, or to
research-grade multi-mode radio SoCs targeting the 400~MHz MICS
band~\cite{Bachmann2015VLSIC}\rev{. To our knowledge, SmartBAN implementations on commercial
SoCs have been limited to our preliminary CC2650 experiments (discussed below)
and the present work.}
Formal methods have been applied to WBAN protocol \emph{verification}
(timed automata, statistical model checking~\cite{BenHamouda2017IINTEC, Touijer2021ComNet}),
but not to resolve specification gaps and inform implementation.
\rev{Prior SmartBAN studies by other groups have been
predominantly simulation-based or analytical (Table~\ref{tab:related}),
although H{\"a}m{\"a}l{\"a}inen~et~al.~\cite{Hamalainen2023ISMICT} confirm
continued interest in standardization. Our previous work evaluated SmartBAN
using TI~CC2650 boards, focusing on initial connection times for a single
Node in isolated~\cite{Hiramatsu2023URSI} and Bluetooth-interfered
environments~\cite{Hiramatsu2024BodyNets}, as well as
a three-Node priority-control scheme~\cite{Ito2024ISMICT}; these
reports did not describe the implementation in detail.
The prototype firmware used in these early studies and in our RSSI
evaluation~\cite{Yoshimi2024GCCE} only supported a hardcoded connection
handshake and a single uplink slot per Node\rev{. Our preliminary
conference report on the nRF54L15 implementation~\cite{Yoshimi2026EMBC}
was limited to an application demonstration without protocol design details}. The present study
develops a new, \rev{standards-aligned} model-driven architecture.}

\begin{table}[!t]
  \caption{Comparison of WBAN and SmartBAN Studies}
  \label{tab:related}
  \centering
  \setlength{\tabcolsep}{2pt}
  \renewcommand{\arraystretch}{1.15}
  \footnotesize
  \begin{tabular}{@{}lllll@{}}
    \toprule
    \textbf{Ref.} & \textbf{Protocol} & \textbf{Approach} & \textbf{Platform} & \makecell[l]{\rev{\textbf{Validation}}\\\rev{\textbf{evidence}}} \\
    \midrule
    \cite{Bachmann2015VLSIC}   & 802.15.6 & MAC+PHY SoC       & Research ASIC  & 400\,MHz,         \\
                               &          & (40\,nm CMOS)     & (IMEC/Fujitsu) & not OTS           \\
    \cite{Sayed2015FPGA}       & 802.15.6 & NB-PHY design     & FPGA       & PHY only          \\
    \cite{Su2017Prototype}     & 802.15.6 & PHY validation    & SDR        & PHY only          \\
    \cite{Boulis2017, Viittala2017}
                               & SmartBAN & Coexistence       & OMNeT++    & Sim., ${\leq}10$\,N  \\
    \cite{Khan2019, Khan2020SmartBAN, KhanFETRO2022}
                               & SmartBAN & MAC optimization  & Simulation & Sim.\ only        \\
    \cite{Ruan2016}            & SmartBAN & Throughput model   & Analytical & ---               \\
    \cite{RamisBibiloni2024}   & SmartBAN & Battery optim.    & Simulation & Sim.\ only        \\
    \cite{DSouza2018ISMICT}    & SmartBAN & BLE comparison    & Simulation & Sim.\ only        \\
    \rev{\cite{Hiramatsu2023URSI, Hiramatsu2024BodyNets, Ito2024ISMICT}}
                               & \rev{SmartBAN} & \rev{Conn.\ time, priority} & \rev{TI CC2650} & \rev{OTA, 1--3\,N} \\
    \rev{\cite{Yoshimi2024GCCE}} & \rev{SmartBAN} & \rev{Application (RSSI)} & \rev{TI CC2650} & \rev{OTA, 6\,N} \\
    \rev{\cite{Yoshimi2026EMBC}} & \rev{SmartBAN} & \rev{Application demo} & \rev{nRF54L15} & \rev{OTA, multi-N} \\
    \midrule
    \textbf{This work}         & \textbf{SmartBAN} & \rev{\textbf{Model-driven}}  & \textbf{Commercial} & \rev{\textbf{OTA, 16\,N,}}   \\
                               &          & \rev{\textbf{impl.}}   & \textbf{SoC}       & \rev{\textbf{25\,h}}   \\
    \bottomrule
  \end{tabular}
  \vspace{1ex}

  \raggedright\footnotesize \rev{OTA: over-the-air RF testbed measurement; Sim.: simulation; N: Nodes; OTS: off-the-shelf.}
\end{table}

\section{Behavioral Modeling}
\label{sec:design}

\subsection{Approach and Methodology}
\label{subsec:approach}
This section describes the hardware-independent behavior of SmartBAN Hub and Node devices as a structured model that addresses the phase-control and connection-lifecycle gaps (Section~\ref{sec:gaps}).

\rev{We} adopt Mealy-type finite automata~\cite{Mealy1955, Hopcroft2006}\rev{, chosen for implementability: each transition produces an explicit output action, so the transition tables guide the firmware dispatch logic without an intermediate synthesis layer.}

We adopt compact notation where $T$ maps (state, event) pairs to (action, next-state) pairs; a single super-state exit ($\tau_{N8}$) follows Harel Statechart notation~\cite{Harel1987, Harel1996} for brevity.
The model abstracts MAC-layer details such as BCH coding, counter thresholds, and scheduling policy, treating them as atomic side-effects within individual transitions; \rev{this separation keeps the state space small.}
Each device model captures the logic governing channel-phase alternation (CCH and DCH) and sub-state transitions in response to received frames and timing events.

\subsection{Behavioral Model of SmartBAN Devices}
\label{subsec:individual_automata}

The SmartBAN protocol behavior is described using two finite automata: one for the Hub device and one for the Node devices. Each is specified as a 5-tuple $(S, E, A, s_0, T)$, where $S$ is the state set, $E$ is the event set, $A$ is the action set, $s_0 \in S$ is the initial state, and $T$ is the transition function.

The Hub coordinates medium access by alternating between Control Channel (CCH) and Data Channel (DCH) phases.

\textbf{Definition 1 (Hub Automaton):}
The Hub automaton $M_H = (S_H, E_H, A_H, s_{0,H}, T_H)$ is defined as:

\begin{enumerate}\raggedright
  \item \textbf{State set}: $S_H = \{CCH\} \cup (\{DCH\} \times \{IDLE, ASGD\})$, where $CCH$ abbreviates the pair $(CCH, -)$ with an irrelevant sub-state, giving three states of uniform product type.
  \item \textbf{Event set}: $E_H = \{Inactive\_End, Slot\_End, CMP\_End,$ $RX\_Uplink, RX\_C\_Req\}$.
  \item \textbf{Action set}: $A_H = \{TX\_C\_Beacon, TX\_D\_Beacon,$ $TX\_I\_Ack, TX\_C\_Assign\}$.
  \item \textbf{Initial state}: $s_{0,H} = (DCH, IDLE)$. The first $CMP\_End$ event triggers $\tau_{H2}$ to transition to CCH.
  \item \textbf{Transition function}: $T_H: S_H \times E_H \rightharpoonup A_H \times S_H$ is a partial function (Table~\ref{table:hub_transitions}); undefined pairs are silently ignored.
\end{enumerate}

\begin{table}[!t]
  \caption{Hub Automaton Transition Rules}
  \label{table:hub_transitions}
  \centering
  \setlength{\tabcolsep}{3pt}
  \renewcommand{\arraystretch}{1.15}
  \begin{tabular}{@{}lllll@{}}
    \toprule
    \textbf{Label} & \textbf{State} $S_H$ & \textbf{Event} $E_H$ & \textbf{Action} $A_H$ & \textbf{Next State} $S_H$ \\
    \midrule
    $\tau_{H1}$    & CCH                  & $Inactive\_End$      & $TX\_D\_Beacon$       & (DCH, IDLE)               \\
    $\tau_{H2}$    & (DCH, IDLE)          & $CMP\_End$           & $TX\_C\_Beacon$       & CCH                       \\
    $\tau_{H3}$    & (DCH, IDLE)          & $RX\_C\_Req$         & $TX\_I\_Ack$          & (DCH, ASGD)               \\
    $\tau_{H4}$    & (DCH, ASGD)          & $Slot\_End$          & $TX\_C\_Assign$       & (DCH, IDLE)               \\
    $\tau_{H5}$    & (DCH, IDLE)          & $RX\_Uplink$         & $TX\_I\_Ack$          & (DCH, IDLE)               \\
    \bottomrule
  \end{tabular}
\end{table}

The Node discovers the Hub via CCH, acquires timing from D-Beacons, and negotiates a connection through the C-Request/C-Assignment exchange. Upon receiving a C-Beacon ($\tau_{N1}$), the Node enters $(DCH, IDLE)$---timing acquired without an active connection---and issues a connection request on the next D-Beacon ($\tau_{N2}$). Once C-Assignment is received ($\tau_{N4}$), the Node enters $(DCH, CONN)$ and transmits its first uplink on the next D-Beacon ($\tau_{N5}$), entering $(DCH, UPLINK)$. \rev{There the model repeats the uplink} on each subsequent D-Beacon ($\tau_{N6}$) until I-Ack ($\tau_{N7}$) returns \rev{the Node} to CONN for the next uplink.

\textbf{Definition 2 (Node Automaton):}
The Node automaton $M_N = (S_N, E_N, A_N, s_{0,N}, T_N)$ is defined as:

\begin{enumerate}\raggedright
  \item \textbf{State set}: $S_N = \{CCH\} \cup (\{DCH\} \times \{IDLE, REQ, CONN, UPLINK\})$, where $CCH$ abbreviates $(CCH, -)$ as in Definition~1, giving five states of uniform product type.
  \item \textbf{Event set}: $E_N = \{Timeout, RX\_C\_Beacon, RX\_D\_Beacon,$ $RX\_I\_Ack, RX\_C\_Assign\}$.
  \item \textbf{Action set}: $A_N = \{TX\_Uplink, TX\_C\_Req, -\}$, where ``$-$'' denotes no output action.
  \item \textbf{Initial state}: $s_{0,N} = CCH$.
  \item \textbf{Transition function}: $T_N: S_N \times E_N \rightharpoonup A_N \times S_N$ is a partial function (Table~\ref{table:node_transitions}); undefined pairs are silently ignored.
\end{enumerate}

\begin{table}[!t]
  \caption{Node Automaton Transition Rules}
  \label{table:node_transitions}
  \centering
  \setlength{\tabcolsep}{3pt}
  \renewcommand{\arraystretch}{1.15}
  \begin{tabular}{@{}lllll@{}}
    \toprule
    \textbf{Label}                    & \textbf{State} $S_N$    & \textbf{Event} $E_N$ & \textbf{Action} $A_N$ & \textbf{Next State} $S_N$ \\
    \midrule
    $\tau_{N1}$                       & CCH                     & $RX\_C\_Beacon$      & -                     & (DCH, IDLE)               \\
    $\tau_{N2}$                       & (DCH, IDLE)             & $RX\_D\_Beacon$      & $TX\_C\_Req$          & (DCH, REQ)                \\
    $\tau_{N3}$\textsuperscript{*}    & (DCH, REQ)              & $RX\_I\_Ack$         & -                     & (DCH, REQ)                \\
    $\tau_{N4}$                       & (DCH, REQ)              & $RX\_C\_Assign$      & -                     & (DCH, CONN)               \\
    $\tau_{N5}$                       & (DCH, CONN)             & $RX\_D\_Beacon$      & $TX\_Uplink$          & (DCH, UPLINK)               \\
    $\tau_{N6}$                       & (DCH, UPLINK)           & $RX\_D\_Beacon$      & $TX\_Uplink$          & (DCH, UPLINK)               \\
    $\tau_{N7}$                       & (DCH, UPLINK)             & $RX\_I\_Ack$         & -                     & (DCH, CONN)               \\
    $\tau_{N8}$\textsuperscript{\dag} & $S_N \setminus \{CCH\}$ & $Timeout$            & -                     & CCH                       \\
    \bottomrule
  \end{tabular}
  \vspace{2pt}
  \raggedright\footnotesize
  \textsuperscript{*}An I-Ack received while in REQ (before C-Assign arrives) is explicitly consumed rather than left undefined\rev{;} the Node continues to await the management-plane C-Assignment response.\\
  \textsuperscript{\dag}$\tau_{N8}$ is a Harel super-state exit from the DCH region. \rev{It models timeout-based disconnection only; the standard's explicit D-Req/D-Res commands (clauses 6.2.6--6.2.7~\cite{ETSI_TS_103_325}) are neither modeled nor implemented.}
\end{table}

\begin{figure}[t]
  \centering
  \includegraphics[width=1.0\columnwidth]{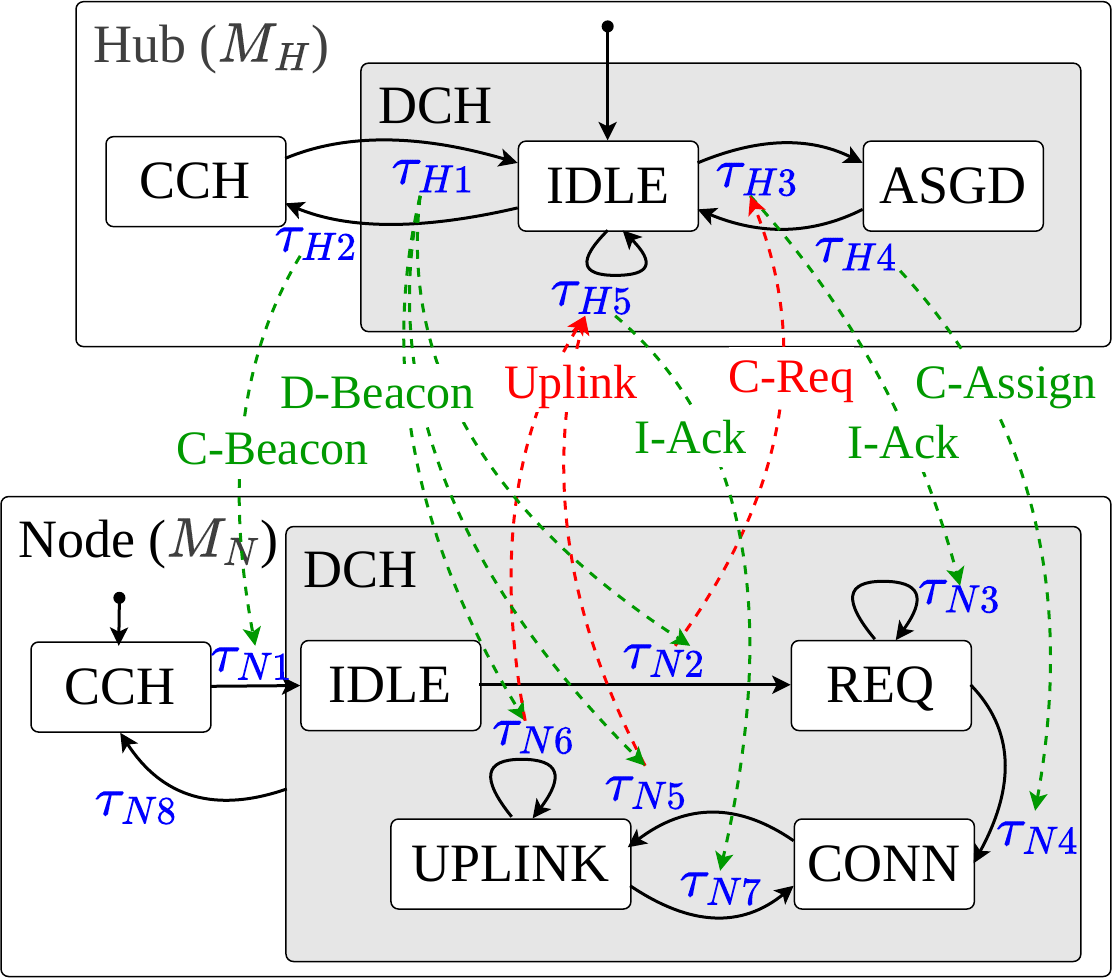}
  \caption{Behavioral model of SmartBAN devices: Visualization of Hub-Node state machine interactions based on Definitions 1 and 2.}
  \label{fig:smartban_simplified_chart}
\end{figure}

Both state machines are single-valued (deterministic in the automata-theoretic sense): for every $(s, e)$ pair in the domain of $T_H$ (respectively $T_N$), Tables~\ref{table:hub_transitions}--\ref{table:node_transitions} specify exactly one $(a, s')$ outcome ($\tau_{N8}$ expands to four flat transitions per the super-state exit semantics, each single-valued); events outside the domain produce no state change.
Fig.~\ref{fig:smartban_simplified_chart} visualizes both models as Harel Statecharts, encapsulating DCH sub-states into super-states. Auxiliary arrows show how one device's output actions trigger the other's input events.

The Hub and Node models interact via broadcast coupling over a shared wireless channel: each transmit action maps to the corresponding receive event in the other device (e.g., $TX\_C\_Beacon \in A_H$ triggers $RX\_C\_Beacon \in E_N$). The $N$ Node replicas execute independently---address-based frame filtering ensures each processes only its own Hub frames---yielding a system comprising one Hub and $N$ independent replicas ($M_H$ with $M_N^{(1)}, \ldots, M_N^{(N)}$).
If a connection request arrives while the Hub is already processing an assignment (state $(DCH, ASGD)$), it falls outside the domain of $T_H$ and is silently ignored; the requesting Node \rev{retries in a later CMP under slotted Aloha contention} (Section~\ref{sec:policies}).

\textbf{Design properties.}
\rev{Three mechanisms are designed to keep both automata progressing:} (i)~the Hub's $CMP\_End$/$Inactive\_End$ timer cycle fires unconditionally, \rev{so the Hub always alternates} between CCH and DCH; (ii)~the Node's $Timeout$ ($\tau_{N8}$) provides a universal escape from any DCH sub-state back to CCH, after which the reconnection path $\tau_{N1} \!\to\! \tau_{N2} \!\to\! \tau_{N4} \!\to\! \tau_{N5}$ re-enters the CONN--UPLINK data-transfer cycle via (DCH, UPLINK); (iii)~\rev{in the model,} the CONN--UPLINK cycle ($\tau_{N5}$--$\tau_{N7}$) \rev{repeats the uplink on each D-Beacon ($\tau_{N6}$) until I-Ack ($\tau_{N7}$) returns the Node to CONN for the next data request.} Because no inter-node state coupling exists, each of $N$ replicas follows the same single-node behavior.

\rev{The evaluated firmware is simpler than property~(iii): it sends a fresh sample
each IBI and never retransmits an unacknowledged frame,
so $\tau_{N6}$ is a design-level provision it does not exercise.}

\textbf{\rev{Extensions.}}
\rev{Operations (2b) and (3)--(5) of Table~\ref{tab:mac_ops} and the D-Req/D-Res exchange of
(6) each map to a bounded extension of $M_H$ or $M_N$: a few states, events, and
transitions tied to existing D-Beacon fields or new management frames. None of these was
implemented; of (6), only timeout-based disconnection detection ($\tau_{N8}$)
is implemented and evaluated.}

\section{Implementation}
\label{sec:implementation}

\subsection{Software Architecture}
\label{sec:software}

\rev{The firmware realizes} the behavioral
model of Section~\ref{sec:design} on a commercial wireless SoC
through the four-layer software architecture of
Fig.~\ref{fig:sw-arch}.
The Application layer handles sensor acquisition; the MAC layer
manages connection state, slot scheduling, and frame assembly;
the PHY layer handles radio configuration and forward error
correction; a hardware abstraction layer (HAL) virtualizes
register-level access and peripheral configuration as the
portability boundary.

\begin{figure}[t]
  \centering
  \begin{tikzpicture}[
    >=latex,
    font=\footnotesize,
    layer/.style={
        draw, thick, rounded corners=1.5pt,
        minimum height=0.85cm,
        align=center, inner sep=2pt,
      },
    submod/.style={
        draw, thick, rounded corners=1pt,
        minimum height=0.65cm,
        align=center, inner sep=2pt, font=\footnotesize,
      },
    annot/.style={font=\footnotesize, text=black!60, align=center},
  ]
  
  \def\lw{6.4}   
  \def\hlw{3.2}  
  
  \node[layer, fill=white, minimum width=\lw cm]
  (app) at (0, 0)
  {Application Layer\\[-1pt]
  {\footnotesize Sensor acquisition, data forwarding}};
  
  \node[layer, fill=black!8, minimum width=\lw cm]
  (mac) at (0, -1.45)
  {MAC Layer\\[-1pt]
  {\footnotesize Behavioral automata (SMF), IBI engine}};
  
  \node[layer, fill=black!16, minimum width=\lw cm]
  (phy) at (0, -2.9)
  {PHY Layer\\[-1pt]
  {\footnotesize BCH codec, CRC}};
  
  \draw[thick, densely dashed, black!50]
  (-\hlw-0.15, -3.65) -- (\hlw+0.15, -3.65)
  node[right, annot] {Portability-\\[-1pt]boundary};
  
  \def\haly{-4.35}
  \def\subw{1.9}
  
  \node[submod, fill=black!28, minimum width=\subw cm]
  (halr) at (-2.25, \haly) {Radio HAL\\[-1pt] GFSK, DMA};
  \node[submod, fill=black!28, minimum width=\subw cm]
  (halt) at (0, \haly) {Timer HAL\\[-1pt] RTC};
  \node[submod, fill=black!28, minimum width=\subw cm]
  (hals) at (2.25, \haly) {Peripheral HAL\\[-1pt] SPI, I\texttwosuperior C};
  
  \draw[thick, decorate,
    decoration={brace, amplitude=3pt}]
  ([xshift=0.3cm]app.north east) -- ([xshift=0.3cm]phy.south east)
  node[midway, right=5pt, annot] {Hardware-\\[-1pt]independent};
  
  \draw[thick, decorate,
    decoration={brace, amplitude=3pt}]
  ([xshift=0.3cm]halr.north east -| phy.east) --
  ([xshift=0.3cm]halr.south east -| phy.east)
  node[midway, right=5pt, annot] {Platform-\\[-1pt]specific};
  
  \node[layer, fill=black!40, minimum width=\lw cm, minimum height=0.65cm,
    font=\footnotesize, text=white]
  (hw) at (0, -5.45)
  {nRF54L15 SoC (Cortex-M33, 2.4\,GHz Radio, GRTC)};
  
  \begin{pgfonlayer}{background}
    \foreach \a/\b in {app/mac, mac/phy} {
        \draw[<->, thick, black!50] (\a.south) -- (\b.north);
      }
    \draw[<->, thick, black!50] ([xshift=-2.25cm]phy.south) -- (halr.north);
    \draw[<->, thick, black!50] (phy.south) -- (halt.north);
    \draw[<->, thick, black!50]
    ([xshift=0.4cm]mac.south) -- ([xshift=0.4cm]halt.north);
    \draw[<->, thick, black!50]
    ([xshift=2.65cm]app.south) -- ([xshift=0.4cm]hals.north);
    \draw[<->, thick, black!50] (halr.south) -- (halr.south |- hw.north);
    \draw[<->, thick, black!50] (halt.south) -- (halt.south |- hw.north);
    \draw[<->, thick, black!50] (hals.south) -- (hals.south |- hw.north);
  \end{pgfonlayer}
  
\end{tikzpicture}
  \caption{Four-layer software architecture deployed on the nRF54L15. \rev{The MAC and PHY layers are
    written against the HAL, so porting to another SoC is expected to require replacing
    only the HAL modules below the portability boundary; no port has been verified yet.}}
  \label{fig:sw-arch}
\end{figure}

\subsection{Behavioral Model Realization}
\label{sec:realization}

Each device state machine of Section~\ref{sec:design} is realized
within the MAC layer, with transitions driven by two event sources:
timer-based slot boundaries and frame reception.
\rev{Every transition of
Tables~\ref{table:hub_transitions}--\ref{table:node_transitions} except
$\tau_{N6}$ (not exercised) has a dispatch path
in the evaluated build, which supports conformance checking. The mapping is
not one-to-one. The Node keeps three connection states (IDLE, REQ, CONN)
and realizes UPLINK as an I-Ack-wait flag. It enters REQ on I-Ack reception,
not on C-Req transmission. Clearing the flag after a two-slot I-Ack timeout
or at the end of the IBI adds an UPLINK-to-CONN edge absent from
Table~\ref{table:node_transitions}.}

\textbf{Timer-driven transitions.}
\rev{The Node derives its slot position in software. It reads the
free-running \qty{1}{\mega\hertz} global real-time counter (GRTC), divides
the time elapsed since the interval origin by the slot duration, and detects
a new slot when the index changes, without a hardware compare interrupt.
It waits for a slot boundary with the RTOS sleep and busy-wait primitives.}
\rev{On D-Beacon reception, the Node re-anchors this interval origin to
the beacon reception timestamp, compensating for inter-IBI clock drift
and filling the slot-level timing gap.}
\rev{The derived slot index drives phase} transitions (e.g.,
$\tau_{H1}$, $\tau_{H2}$, $\tau_{N8}$).
At the DCH/CCH boundary, the radio
is reconfigured from the data frequency to the control frequency.
Because a single radio transceiver handles both C-Beacon and D-Beacon
traffic, the two frame types are inherently separated by the
phase structure of the automaton---no additional arbitration is required.

\textbf{Event-driven transitions.}
The PHY layer validates the cyclic redundancy check (CRC) and performs
BCH decoding on each received frame.
For valid frames, the MAC layer dispatches to the appropriate handler
based on the frame type and subtype fields extracted from the
Frame Control header.
Each handler references the device's current state to select the
appropriate transition (e.g., $\tau_{H3}$, $\tau_{H5}$ for the Hub;
\rev{$\tau_{N1}$, $\tau_{N3}$, $\tau_{N4}$, and $\tau_{N7}$} for the Node)\rev{. After D-Beacon reception, the $\tau_{N2}$ and
$\tau_{N5}$ transmissions fire in the next CMP and the assigned slot,
respectively}.
\rev{The I-Ack frame is assembled on demand within the frame handler,
achieving the sub-millisecond TX/RX turnaround required for
acknowledged uplink reception.}

\subsection{Implementation Details}
\label{sec:impl-details}

\textbf{PHY layer.}
Following the SmartBAN PHY specification~\cite{ETSI_TS_103_326},
the radio operates in proprietary GFSK mode
($h{=}0.5$, $\mathrm{BT}{=}0.5$, \qty{1}{\mega\bit\per\second})
with a 32-bit preamble, bypassing the SoC's BLE stack.
\rev{When enabled,} BCH(127,113,$t{=}2$) forward error correction protects all frames, using
pre-computed $\mathrm{GF}(2^{7})$ lookup tables (\qty{896}{\byte} for
exponentiation, logarithm, and parity tables).
\rev{In this implementation, the control and data channels are fixed at build time.
The Hub transmits C-Beacons on control channel~0, and the Node listens on that
channel only; scanning the three control channels is not implemented.
A single data channel carries all scheduled traffic.}

\textbf{MAC layer.}
Both Hub and Node maintain a \emph{Node Table}---a per-connection context
structure that tracks the connection sub-state defined in the
transition tables---so that the set of connected Nodes is determined at
runtime rather than fixed at compile time.
All timing and configuration values (slot width, inter-frame spacing,
IBI, etc.) are parameterized as compile-time constants.

\subsection{Scheduling and Timeout Policies}
\label{sec:policies}

\textbf{Slot allocation.}
The Hub uses first-fit sequential allocation from slot~1, assigning
$N_{\text{sap}}$ consecutive free slots to minimize allocation latency
for up to $N_{\max} = 16$ nodes.

\textbf{\rev{Connection setup.}}
\rev{An unconnected Node sends its C-Req in the CMP by slotted Aloha. The Hub
acknowledges the C-Req with an I-Ack and sends the C-Assignment in a later CMP
slot. A C-Req that receives no I-Ack is retried: the transmit probability starts
at $1/8$ and drops to $1/16$ after two consecutive failures, with no retry
limit. The Node does not acknowledge the C-Assignment, so the Hub treats the
Node's first scheduled uplink as confirmation and only then marks it connected.}

\textbf{Timeout and recovery.}
\rev{Three timeouts, counted in IBIs (Table~\ref{tab:timing}), send a Node back
to the CCH to await a C-Beacon ($\tau_{N8}$): no C-Assignment within $N_{\text{cass}}$
IBIs after the C-Req is acknowledged, no D-Beacon for $N_{\text{miss}}$ consecutive IBIs, or no
I-Ack for $N_{\text{iack}}$ consecutive IBIs. The Hub likewise releases a
Node's slots after $N_{\text{miss}}$ IBIs without an uplink, a policy outside
the Hub automaton. Each value leaves a margin over normal operation: a C-Req
and its C-Assignment fit in the CMP of a single IBI, and $N_{\text{miss}}$ was
tuned in preliminary multi-node runs so that occasional beacon losses do not
disconnect a Node. $N_{\text{iack}}$ is smaller than $N_{\text{miss}}$, so a
Node whose uplinks stop reaching the Hub gives up before the Hub releases its
slot to another Node.}

\textbf{\rev{Uplink handling.}}
\rev{In the evaluated firmware, the Node sends a fresh sensor sample in its
slot every IBI, whether or not an I-Ack arrived for the previous one. An
unacknowledged frame is replaced by the next sample rather than retransmitted.
A bounded-retry mode for discrete
events exists but is disabled in every evaluated build; the only
retransmissions are C-Req retries.}

\section{Experimental Evaluation}
\label{sec:experiments}

Experiments target operations~\rev{(1), (2a), and the timeout path of~(6)} of
Table~\ref{tab:mac_ops}, \rev{exercising the transitions of $M_H$ and
$M_N$ that these paths reach}.

\begin{table}[b]
  \centering
  \caption{Timing and Configuration Parameters}
  \label{tab:timing}
  \setlength{\tabcolsep}{3pt}
  \renewcommand{\arraystretch}{1.1}
  \begin{tabular}{@{}llll@{}}
    \toprule
    \textbf{Parameter} & \textbf{Symbol}    & \textbf{Value}           & \textbf{Note}                                 \\
    \midrule
    \multicolumn{4}{@{}l}{\textit{Standard constants}}                                                                 \\
    Min.\ time unit    & $T_{\min}$         & \qty{625}{\micro\second} & Fixed                                         \\
    Slot length        & $L_{\text{slot}}$  & 16                       & One of six standard options                    \\
    Max nodes          & $N_{\max}$         & 16                       & Fixed                                         \\
    \midrule
    \multicolumn{4}{@{}l}{\textit{Design parameters}}                                                                  \\
    Inter-frame space  & $T_{\text{IFS}}$   & \qty{150}{\micro\second} & Fixed (TIFS)                                  \\
    Slots per node     & $N_{\text{sap}}$   & 1                        & Configurable                                  \\
    CMP slots          & $N_{\text{cmp}}$   & 2                        & Configurable                                  \\
    Inactive slots     & $N_{\text{inact}}$ & 3                        & Configurable                                  \\
    \rev{C-Assign.\ wait}       & \rev{$N_{\text{cass}}$} & \rev{3 IBIs}  & \rev{Configurable}                      \\
    \rev{D-Beacon miss limit}   & \rev{$N_{\text{miss}}$} & \rev{10 IBIs} & \rev{Configurable}                      \\
    \rev{I-Ack miss limit}      & \rev{$N_{\text{iack}}$} & \rev{5 IBIs}  & \rev{Configurable}                      \\
    \midrule
    \multicolumn{4}{@{}l}{\textit{Derived values}}                                                                     \\
    Slot duration      & $T_S$              & \qty{10}{\milli\second}  & $L_{\text{slot}} \times T_{\min}$             \\
    SAP slots          & $N_S$              & 16                       & $N_{\max} \times N_{\text{sap}}$              \\
    Slots per IBI      & $L_D$              & 22                       & $1 + N_S + N_{\text{cmp}} + N_{\text{inact}}$ \\
    IBI duration       & $T_D$              & \qty{220}{\milli\second} & $L_D \times T_S$                              \\
    \bottomrule
  \end{tabular}
\end{table}

\subsection{Experimental Setup}
\label{sec:exp:setup}

\begin{figure}[t]
  \centering
  \includegraphics[width=0.8\columnwidth]{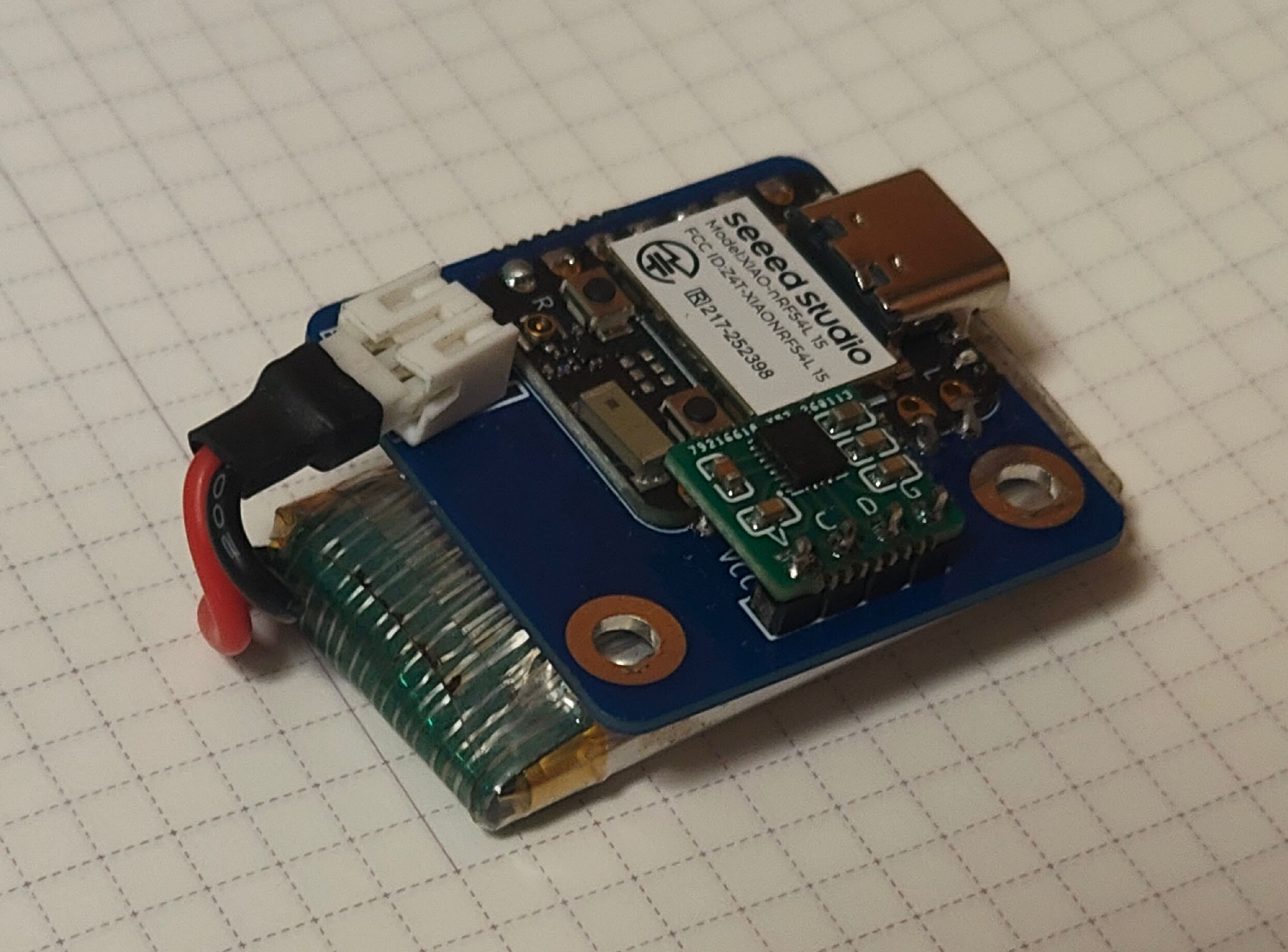}
  \caption{Sensor node: XIAO nRF54L15~\cite{XIAO54L15Doc}
    ($21 \times 17.8$\,mm, ${\sim}$USD\,6) with Nordic
    nRF54L15~\cite{NordicnRF54L15} SoC (Arm Cortex-M33,
    \qty{128}{\mega\hertz}, \qty{1.5}{\mega\byte} Flash,
    \qty{256}{\kilo\byte} RAM) running
    Zephyr RTOS v4.2~\cite{ZephyrProject} /
    nRF Connect SDK~\cite{NordicNCSSDK} v3.2.1,
    custom PCB carrying an LSM9DS1 IMU (I\textsuperscript{2}C),
    and battery holder.
    Firmware footprint: \qty{59}{\kilo\byte}/\qty{63}{\kilo\byte}
    Flash (Hub/Node), ${\sim}\qty{35}{\kilo\byte}$ RAM each
    (4\% / 14\% of available capacity).}
  \label{fig:node-photo}
\end{figure}

Fig.~\ref{fig:node-photo} shows the sensor node used in all experiments \rev{except
current profiling, which used a bare XIAO nRF54L15 without the carrier PCB}.
\rev{The Hub uses the same hardware with its own build (Fig.~\ref{fig:node-photo}) and
connects to a host PC via UART for logging.}

All experiments were performed indoors in a shared laboratory with typical ambient
Wi-Fi and Bluetooth activity, at \qty{1}{\metre} Hub--Node distance \rev{except the interference
runs (Section~\ref{sec:exp:interference}), whose geometry raised the median Node RSSI by
\qty{12}{dB}}.
Table~\ref{tab:timing} summarizes the timing and configuration
parameters common to all experiments.
\rev{One uplink slot per Node corresponds to the per-IBI sensor data batch.
Two CMP slots allow a connection request and its assignment within a single interval.
Three inactive slots (\qty{30}{\milli\second}) reserve time for the Hub to output its
per-IBI event log over UART before the next IBI, so that UART DMA does not overlap radio
reception.
The \qty{10}{\milli\second} slot length, a standard option, provides sufficient headroom
for the frame, its acknowledgment, and guard intervals.
These values represent a baseline design choice; Fig.~\ref{fig:iavg_vs_ibi} evaluates
how the energy cost varies with the IBI.}
A Nordic PPK2~\cite{NordicPPK2} in source mode supplied \qty{3.8}{\volt}
via the BAT+ pad \rev{(nominal \qty{100}{kS/s}, \qtyrange{78.8}{81.5}{kS/s} achieved;}
\qty{0.2}{\micro\ampere} resolution, $\pm$5\% systematic accuracy)
for board-level current measurement.
Only the current-profiling firmware disables non-protocol peripherals
to isolate the energy footprint (\rev{five} GPIO channels mark protocol
states for per-IBI accounting); all other experiments use the full
application firmware with the IMU sampled \rev{every \qty{8}{\milli\second}}.
\rev{Timing and delivery results come from a \qty{25}{\hour} run of
the 16-Node network, a 1-h run without BCH, three 1-h captures taken back to back on the
same day (SmartBAN, and BLE with 16 and with one peripheral; Fig.~\ref{fig:iat_hist},
Table~\ref{tab:ble_comparison}), and the interference runs.
Current captures last \qty{900}{\second}. The current captures and the same-day captures
discard their first \qty{120}{\second}, and the interference runs their first \qty{30}{\second}.}

\subsection{Behavioral Conformance}
\label{sec:exp:conformance}

\rev{We assess behavioral conformance by comparing observed
state transitions against the behavioral model.}
\rev{The Hub logs every transmitted and received frame with a hardware timestamp,
sequence number, and Node identifier. Because device states are not directly logged,
we infer state transitions from the recorded event sequence.
Frame exchanges directly evidence all five Hub transitions (Table~\ref{table:hub_transitions}).
Hub-observable consequences indirectly verify seven of the eight Node transitions
(Table~\ref{table:node_transitions}).
The remaining transition, $\tau_{N6}$, is excluded from coverage because the evaluated
build never retransmits an uplink and so does not exercise it.
Observed per-transition counts range from 1 (for $\tau_{N8}$, via disconnection detection)
to $6.55 \times 10^{6}$ (for $\tau_{H5}$).}

\textbf{Connection lifecycle.}
One disconnection event triggered $\tau_{N8}$ via the timeout
mechanism of Section~\ref{sec:policies}; the Node returned to
CCH and completed the full reconnection sequence
($\tau_{N1} \!\to\! \tau_{N2} \!\to\! \tau_{N4} \!\to\! \tau_{N5}$)
as modeled.

\begin{figure}[t]
  \centering
  \includegraphics[width=\columnwidth]{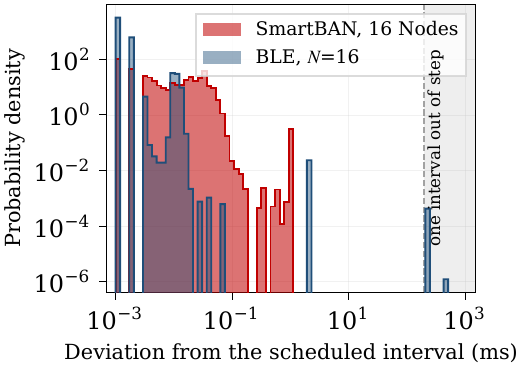}
  \caption{\rev{Arrival timing deviation between consecutively numbered frames in the
  same-day captures of 16-Node SmartBAN and BLE (definitions in Table~\ref{tab:ble_comparison}).
  The shaded band holds BLE deferrals, each displacing two intervals.
  Exact zeros are drawn at \qty{1}{\micro\second}.}}
  \label{fig:iat_hist}
\end{figure}

\textbf{Timing precision.}
\rev{Because the Hub's D-Beacon interval itself varies by about \qty{0.9}{\milli\second},
we measure each uplink's arrival relative to the D-Beacon of its IBI.
Over the 25-h run, 13 of the 16 Nodes keep a standard deviation of
\qtyrange{8.5}{12.0}{\micro\second}.
The other three occasionally arrive about \qty{0.93}{\milli\second} early
(4.6--7.3\% of IBIs) for a reason we did not identify; this did not recur in the
later captures (Fig.~\ref{fig:iat_hist}, Table~\ref{tab:interference}).}

\subsection{\texorpdfstring{\rev{Standards-Aligned Communication}}{Standards-Aligned Communication}}
\label{sec:exp:comm}

\textbf{Communication quality.}
\rev{Over the 1-h window of the 25-h run, 8 frames were lost (packet delivery ratio, PDR,
99.997\%).
PHY decode failures were rare with and without BCH coding (about two per \num{100000}
receptions in each 1-h run); under these line-of-sight conditions the high link margin
left no errors to correct, so BCH coding gain was not exercised.}

\begin{figure*}[t]
  \centering
  \includegraphics[width=\linewidth]{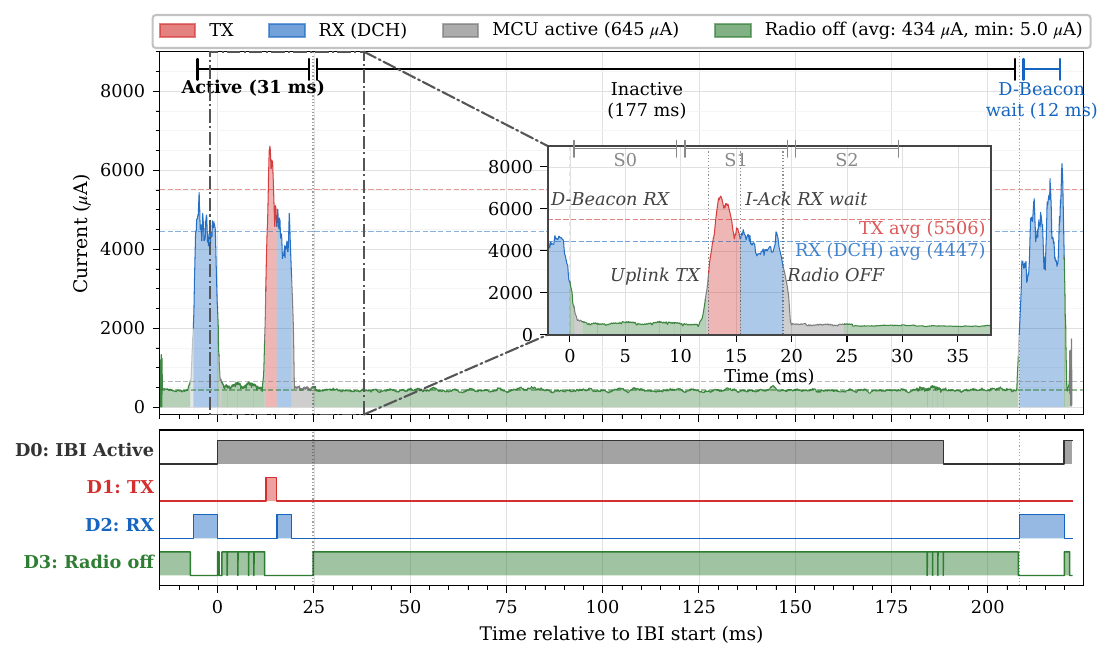}
  \caption{\rev{Measured current profile of one IBI
    ($T_D = \qty{220}{\milli\second}$, Node), from the capture of Table~\ref{tab:current}.
    Colors follow the protocol states that firmware GPIO markers define (lower panel).
    Horizontal dashed lines show per-state averages for this interval, and vertical dotted lines mark
    phase transitions. The inset magnifies the active phase; its faint dashed lines mark
    \qty{10}{\milli\second} slot boundaries.}}
  \label{fig:current_waveform}
\end{figure*}

\textbf{Energy efficiency.}\label{sec:exp:power}
Fig.~\ref{fig:current_waveform} and Table~\ref{tab:current} show the
measured current profile for one IBI.
\rev{The radio is off for nine tenths of the interval, yet this state carries more than half of
the average current; the radio operations themselves draw about $1.4\times$ their datasheet
values (Table~\ref{tab:current}; not a like-for-like ratio, as its note states).
Between events the current falls to about \qty{5}{\micro\ampere} (Fig.~\ref{fig:current_waveform}),
far below the radio-off average, which therefore reflects activity rather than a floor.
The board-level peak of \qty{37.4}{\milli\ampere} constrains supply and cell selection, while
the \qty{0.73}{\milli\ampere} average sets operating life: about 26~days on a
\qty{450}{\milli\ampere\hour} cell at the \qty{220}{\milli\second} IBI.}

\begin{figure}[t]
  \centering
  \includegraphics[width=\columnwidth]{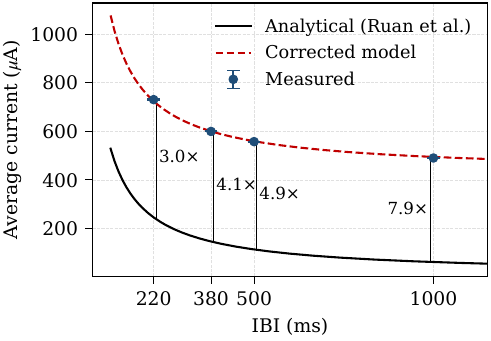}
  \caption{Average current $I_{\text{avg}}$ versus IBI.
    Markers: measured values\rev{. Error bars for the 95\% confidence interval (\qtyrange{1.0}{2.3}{\micro\ampere})
    are smaller than the markers.}
    Solid line: analytical model of
    Ruan~et~al.~\cite{Ruan2018JBHI} using nominal current values.
    Dashed line: corrected model using measured state currents
    from Table~\ref{tab:current}.}
  \label{fig:iavg_vs_ibi}
\end{figure}

\begin{table}[b]
  \centering
  \begin{threeparttable}
    \caption{\rev{Current Consumption and Timing Breakdown ($T_D = \qty{220}{\milli\second}$)}}
    \label{tab:current}
    \setlength{\tabcolsep}{3pt}
    \renewcommand{\arraystretch}{1.15}
    \begin{tabular}{@{}lrrrrr@{}}
      \toprule
      \textbf{State}
       & \rev{\textbf{Duration}}
       & \rev{\textbf{Duty}}
       & \rev{\textbf{Measured}}
       & \textbf{Nominal}
       & \textbf{Ratio} \\
       & \rev{(\unit{\milli\second})}
       & \rev{(\%)}
       & \rev{(Mean $\pm$ Std)}
       & \rev{(\unit{\milli\ampere})}
       & \\
      \midrule
      TX (\qty{0}{dBm})
       & \rev{2.18}
       & \rev{0.99}
       & \rev{$6.82 \pm 4.53$~\unit{\milli\ampere}}
       & 4.8
       & \rev{$1.42\times$} \\
      RX\textsuperscript{a}
       & \rev{12.22}
       & \rev{5.56}
       & \rev{$4.59 \pm 2.72$~\unit{\milli\ampere}}
       & 3.4
       & \rev{$1.35\times$} \\
      Radio off\textsuperscript{b}
       & \rev{199.58}
       & \rev{90.72}
       & \rev{$0.43 \pm 0.41$~\unit{\milli\ampere}}
       & ---
       & --- \\
      \rev{MCU active\textsuperscript{c}}
       & \rev{6.02}
       & \rev{2.73}
       & \rev{$0.68 \pm 0.89$~\unit{\milli\ampere}}
       & \rev{---}
       & \rev{---} \\
      \midrule
      Peak (observed)
       & ---
       & ---
       & \rev{37.4~\unit{\milli\ampere}}
       & ---
       & --- \\
      \bottomrule
    \end{tabular}

    \begin{tablenotes}
      \footnotesize
      \item \rev{Board-level currents of a Node, with states set by the GPIO markers
      (Section~\ref{sec:exp:setup}).
      Nominal values are SoC datasheet figures at their own supply conditions, so Ratio
      is not a like-for-like comparison.}
      \item[a] \rev{Listening and frame decoding.}
      \item[b] \rev{Radio transceiver disabled.}
      \item[c] \rev{No TX, RX, or radio-off marker set.}
    \end{tablenotes}
  \end{threeparttable}
\end{table}

\subsection{Long-Term Stability}
\label{sec:exp:stability}

\rev{The \qty{25}{\hour} continuous test (BCH enabled) ran to completion with all 16 Nodes
transmitting live IMU data. Memory and buffer usage were not instrumented; we judge stability
from delivery, from the per-Node slot alignment of Section~\ref{sec:exp:conformance} over the
whole run, and from the single disconnection below.}

Aggregate PDR exceeded 99.99\% \rev{(every Node above 99.98\%), and
per-node PDR showed no correlation with RSSI.}
Only one disconnection occurred\rev{, at Node~2, the Node with the weakest link.
Its uplinks stopped for \qty{2.4}{\second}}, consistent with \rev{the Hub's} timeout detection
($N_{\text{miss}} \times T_D \approx \qty{2.2}{\second}$)
\rev{followed by reconnection within \qty{264}{\milli\second} (1.2~IBIs).
Its RSSI returned to its mean afterwards, consistent with a transient disturbance;
RF channel activity was not logged, so the trigger cannot be determined.}

\subsection{\texorpdfstring{\rev{Operation Under Interference}}{Operation Under Interference}}
\label{sec:exp:interference}

\rev{SmartBAN's control channels (CCH~0--2) share center frequencies with BLE advertising
channels~37--39, and its data channels can coincide with active BLE connections. We placed nRF54L15 BLE interferers near the 16-Node network in two controlled experiments
(Table~\ref{tab:interference}).

\textbf{1) Impact on connectivity.}
Four BLE advertisers on channels~37--39, at up to \qty[retain-explicit-plus]{+8}{dBm}, raised
the peak power on channel~38 by about \qty{30}{dB}.
No uplink was lost, and every Node joined in each of three trials with the advertisers
active, as in three interferer-free trials.

\textbf{2) Impact on data transmission.}
The advertisers never transmit on a SmartBAN data channel, so these runs placed no
co-channel energy on the uplinks. We therefore pinned one or two BLE links at
\qty[retain-explicit-plus]{+8}{dBm} to channels~35/36, placing co-channel energy on DCH~36.
SmartBAN lost none of the \num{58902} scheduled uplinks, which bounds the uplink loss rate
below 50.9~ppm (one-sided 95\%, Clopper--Pearson), and per-Node timing $\sigma$ stayed at the
interferer-free level.
Interferer power at the Hub and its distance were not recorded, and the uplinks were about
\qty{12}{dB} stronger than in the 25-h run, so zero loss shows only that interference stayed
below the disruptive level in this geometry; no degradation threshold was established.}

\begin{table}[t]
  \centering
  \caption{\rev{Controlled RF Interference Evaluation}}
  \label{tab:interference}
  \setlength{\tabcolsep}{2pt}
  \renewcommand{\arraystretch}{1.12}
  \rev{\begin{tabular}{@{}lll@{}}
    \toprule
    \textbf{Metric / Parameter}
      & \textbf{Control Plane}
      & \textbf{Data Plane} \\
    \midrule
    Target channel
      & Control (CCH~0)
      & Data (DCH~36) \\
    Applied interferer
      & 4 BLE adv.\ (ch~37--39)
      & 1--2 BLE links \\
    Interferer TX power
      & \qty{-40}{} / \qty[retain-explicit-plus]{+8}{dBm}
      & \qty[retain-explicit-plus]{+8}{dBm} (max) \\
    Peak intf.\ power\textsuperscript{a}
      & \qty{-38.0}{} $\to$ \qty{-8.4}{dBm}
      & \qty{-40.2}{} $\to$ \qty{-7.5}{dBm} \\
    Power increase ($\Delta_{\text{peak}}$)
      & \qty[retain-explicit-plus]{+29.6}{dB}
      & \qty[retain-explicit-plus]{+32.7}{dB}\textsuperscript{b} \\
    \midrule
    Joins, \qty[retain-explicit-plus]{+8}{dBm} active
      & 48/48 (100\%)
      & --- \\
    Uplinks lost
      & 0
      & 0 \\
    Per-Node $\sigma$\textsuperscript{c}
      & \qtyrange{8.2}{14.7}{\micro\second}
      & \qtyrange{7.4}{15.1}{\micro\second} \\
    \bottomrule
  \end{tabular}}
  \vspace{4pt}
  \raggedright\footnotesize
  \rev{\textsuperscript{a}\,Clean baseline $\to$ interferers active; control plane on channel~38 to isolate interferer emissions from SmartBAN traffic, data plane from the pinned run (\qty{7.5}{\milli\second} interval).\\
  \textsuperscript{b}\,In mean terms, \qty[retain-explicit-plus]{+23.5}{dB} (\qty{-63.1}{} to \qty{-39.6}{dBm}).\\
  \textsuperscript{c}\,Uplink offset from the D-Beacon, all 16 Nodes; same-day interferer-free baseline \qtyrange{8.8}{14.4}{\micro\second}.}
\end{table}

\section{Discussion}
\label{sec:discussion}

\subsection{\texorpdfstring{\rev{Energy Model Validation}}{Energy Model Validation}}
\label{sec:disc:model}

We assess the validity of the Ruan~et~al.\ analytical model against
the measured current profile
(Fig.~\ref{fig:iavg_vs_ibi}, Table~\ref{tab:current}).
Following Ruan~et~al.~\cite{Ruan2018JBHI}, the average current is
evaluated using a three-state weighted average:
\begin{equation}
  I_{\text{avg}}^{(a)} = I_{\text{TX}} \cdot d_{\text{TX}}
    + I_{\text{RX}} \cdot d_{\text{RX}}
    + I_{\text{idle}} \cdot (1 - d_{\text{TX}} - d_{\text{RX}}),
  \label{eq:power_model}
\end{equation}
where $d_{\text{TX}}$ and $d_{\text{RX}}$ are the TX and RX duty cycles
(Table~\ref{tab:current}: \rev{$d_{\text{TX}} = 0.99\%$,
$d_{\text{RX}} = 5.56\%$}).
Using the nominal values from~\cite{NordicnRF54L15}
($I_{\text{TX}} = \qty{4.8}{\milli\ampere}$,
$I_{\text{RX}} = \qty{3.4}{\milli\ampere}$,
$I_{\text{idle}} = \qty{0.01}{\milli\ampere}$),
Fig.~\ref{fig:iavg_vs_ibi} shows that the model underestimates
measured $I_{\text{avg}}$ by \rev{$3.0$--$7.9\times$}.
Substituting the measured state currents of Table~\ref{tab:current}
into~\eqref{eq:power_model} reduces the maximum absolute error to
\rev{0.9\%}.
\rev{The state currents come from the \qty{220}{\milli\second} run, yet they
predict the other three intervals just as closely, so the form of~\eqref{eq:power_model}
is consistent with the measurement.

The gap to the nominal model lies mainly in the radio-off state.
The Node draws \qty{0.43}{\milli\ampere} with the radio off instead of the
datasheet's \qty{0.01}{\milli\ampere}, and the radio is off for most of each
interval (Table~\ref{tab:current}).
At \qty{220}{\milli\second}, the radio-off state accounts for 78\% of the
excess over the nominal model, and the other three states for the rest.
With the active durations of Table~\ref{tab:current} at every IBI, the active
terms of~\eqref{eq:power_model} fall as $1/\text{IBI}$ while the idle term stays.
The measured current therefore levels off near the radio-off current, whereas the
nominal model, with almost no idle term, keeps falling; this is why the
underestimate grows with the IBI (Fig.~\ref{fig:iavg_vs_ibi}).
The same board draws a median of \qty{1.1}{\micro\ampere} with a radio-free
build, so the hardware does not set the radio-off current; we attribute it to
background tasks and slot-timing wakeups, an implementation-level overhead
that we expect to be addressable without modifying the protocol.
For battery-life estimates, \eqref{eq:power_model} therefore needs state
currents measured on the implementation, above all $I_{\text{idle}}$; with
datasheet values it would overstate this implementation's operating life by the
same factor.}

\subsection{BLE Comparison}
\label{sec:disc:ble}

\rev{To characterize both protocols,
we compared SmartBAN with BLE on the identical SoC platform}
(nRF54L15, Zephyr BLE stack) with matched 1M~PHY,
\rev{connection interval (CI)}${} = \qty{220}{\milli\second}$, and slave latency~0
(Table~\ref{tab:ble_comparison}).

\begin{table}[t]
  \centering
  \caption{\rev{Same-SoC Protocol Comparison: SmartBAN vs. BLE (nRF54L15)}}
  \label{tab:ble_comparison}
  \setlength{\tabcolsep}{4pt}
  \renewcommand{\arraystretch}{1.15}
  \begin{tabular}{@{}lrrr@{}}
    \toprule
    \textbf{Metric}
     & \textbf{SmartBAN}
     & \multicolumn{2}{c}{\rev{\textbf{BLE}}} \\
    \cmidrule(lr){3-4}
    \rev{Number of Nodes ($N$)}
     & \rev{16}          & \rev{16}         & \rev{1}          \\
    \midrule
    \multicolumn{4}{@{}l}{\rev{\textbf{Energy Efficiency (220-ms interval)}}} \\
    $I_{\text{avg}}$ (\unit{\micro\ampere})\textsuperscript{\rev{a}}
     & \rev{731}         & \rev{327}        & \rev{327}        \\
    \rev{Battery life (days)}\textsuperscript{\rev{b}}
     & \rev{25.7}        & \rev{57.3}       & \rev{57.3}       \\
    \midrule
    \multicolumn{4}{@{}l}{\rev{\textbf{Timing and Delivery}}} \\
    \rev{Frame lost (\%)}
     & \rev{0.0069}      & \rev{0} & \rev{0} \\
    \rev{Deferred (\%)}
     & \rev{0}  & \rev{0.905}      & \rev{0.724}      \\
    \rev{Median deviation (\unit{\micro\second})}
     & \rev{30}          & \rev{1} & \rev{1} \\
    \rev{$P_{99}$ deviation (\unit{\milli\second})}
     & \rev{1.03} & \rev{217.9}     & \rev{217.9}      \\
    \rev{Max deviation (\unit{\milli\second})}
     & \rev{1.08} & \rev{440.0}     & \rev{220.0}      \\
    \bottomrule
  \end{tabular}

  \vspace{2pt}
  \raggedright\footnotesize
  \rev{193~B payload, \qty{0}{dBm} TX, 1M~PHY; timing and loss from the same-day \qty{1}{\hour} captures (Fig.~\ref{fig:iat_hist}).}\\
  \textsuperscript{\rev{a}}\,\rev{Measured as described in Section~\ref{sec:exp:setup}; independent of network scale within 0.6\%.}\\
  \textsuperscript{\rev{b}}\,\rev{Projected for a 450~mAh LiPo cell in steady-state operation.}\\
  \rev{Deviation: $|\Delta t - \qty{220}{\milli\second}|$ between consecutively numbered frames; ``deferred'': ${\geq}0.9$ interval late. Timestamps: SmartBAN, Hub GRTC at frame arrival; BLE, Central log time.}
\end{table}

\textbf{Timing.}
\rev{The evaluated SmartBAN build does not retransmit uplinks, whereas BLE's link
layer retransmits unacknowledged packets.
On the same bench, SmartBAN lost 0.0069\% of frames and deferred none, while
BLE lost none but deferred 0.9\% by one or two intervals
(Table~\ref{tab:ble_comparison}).
Consecutively numbered frames from each SmartBAN Node arrived
\qty{220}{\milli\second} apart to within \qty{1.08}{\milli\second}.
This spread includes the Hub's beacon-interval variation, about
\qty{1}{\milli\second} in the same-day capture, and SmartBAN's median
deviation includes the offset of the Hub's median beacon period from
\qty{220}{\milli\second}; neither is a measure of slot error.
BLE's lower median deviation reflects its timestamp source: the Central log
time follows the Central's own connection-event clock.}

\rev{Under these link policies, SmartBAN's departures from the schedule appear as lost frames and BLE's as late frames.
The observed SmartBAN range suggests that applications can size their
buffers from the TDMA schedule with a millisecond-scale margin, although
the range is not a guaranteed bound.
Which behavior an application prefers depends on whether it tolerates a
rare gap or an interval-scale delay; with uplink retransmission enabled, or
at a different interval, the picture could differ.}

\textbf{Energy \rev{and battery life}.}
\rev{SmartBAN draws $2.2\times$ BLE's average current, so on the same cell its
projected operating life is less than half of BLE's (Table~\ref{tab:ble_comparison}).
Table~\ref{tab:current} decomposes SmartBAN's current by state.
The BLE current was not resolved by state, so this decomposition does not apportion the gap between the two protocols.
Two features of the SmartBAN firmware are nonetheless candidate causes.
Its radio-off state alone contributes \qty{390}{\micro\ampere} to the average,
more than BLE's whole average.
In addition, the receiver turns on one slot before each expected D-Beacon, so
most of the receive time per interval (Table~\ref{tab:current}) is spent waiting
for the beacon rather than receiving it, about \qty{200}{\micro\ampere} of the
average current.
Both projected lives exceed three weeks, but they cover the protocol firmware
alone: the current captures ran on a bare module without the sensor, so a
sensing node will run for less.}

\textbf{Channel diversity.}
SmartBAN operates on a single fixed data channel, whereas BLE
employs \rev{frequency hopping} across 37~data channels;
in interference-rich environments, this asymmetry would favor
BLE's PDR.
\rev{SmartBAN defines a data channel migration procedure (operation~(5) in Table~\ref{tab:mac_ops}) but not its trigger conditions or channel-quality metrics, and our firmware does not implement it.
Co-channel BLE traffic caused no uplink loss (Table~\ref{tab:interference}), so the interference level at which the fixed data channel would need migration remains unknown.}

\subsection{Limitations and Future Work}
\label{sec:disc:limits}

\textbf{Experimental scope \rev{and on-body evaluation}.}
\rev{Our evaluations were conducted on a static bench-top testbed using a single SoC
family (nRF54L15).
More importantly, bench testing does not capture the dynamic body shadowing,
multipath fading, and postural changes inherent to wearable deployments.
Conducting on-body trials under user mobility and experimentally verifying cross-platform
portability across different wireless SoCs represent our primary experimental priorities.}

\textbf{\rev{Protocol scope and functional extensions.}}
\rev{The operations of Table~\ref{tab:mac_ops} that this work does not implement, such as
contention-period uplink~(2b), data channel migration~(5), and the explicit disconnection
exchange of~(6), would add states and transitions to the behavioral model.
We plan to implement them, together with security and encryption, priority-based scheduling
to improve on the baseline first-fit slot allocation, and optional
multi-hop relay~\cite{ETSI_TS_103_805}.}

\section{Conclusion}
\label{sec:conclusion}

\rev{We implemented SmartBAN initial connection, scheduled uplink, and
timeout-based disconnection detection on a commercial wireless SoC
(nRF54L15) in a four-layer software architecture.
Structured behavioral modeling provided Hub and Node Mealy automata as a
design blueprint for the phase-control and connection-lifecycle behavior
that the standard leaves unspecified.
Every transition but one has a firmware dispatch path.
In a 25-h run with 16 Nodes, the Hub's frame log covered every transition
with a dispatch path, nearly all frames were delivered, and timeout-triggered
reconnection followed the single observed disconnection.
Separate current measurements show that an analytical model with datasheet
currents underestimates the average current, mostly in the radio-off state,
where we attribute the current to implementation-level overhead.}

\rev{In a same-SoC comparison with BLE, our SmartBAN build, configured to send
a fresh sample every interval instead of retransmitting, kept each Node's frame
spacing within \qty{1.1}{\milli\second} of the interval but lost some
frames, whereas BLE lost none but deferred some by one or two intervals.
BLE drew less than half SmartBAN's average current, although both are
projected to run for weeks on a small battery.
Future work includes on-body trials under user mobility, portability tests on
other SoCs, and implementation of the remaining operations as extensions of
the automata.}

\bibliographystyle{IEEEtran}
\bibliography{ref}

\end{document}